\documentclass[prodmode,acmtomacs]{acmsmall}
\usepackage{graphicx}
\usepackage{hyperref}
\usepackage{mathtools}
\usepackage{xspace}
\usepackage{prodint}
\usepackage{algorithmic}

\doi{0000001.0000001}

\issn{1234-56789}

\newcommand{\ctde}{\textsc{ctde}\xspace}
\newcommand{\ssa}{\textsc{ssa}\xspace}

\newcommand{\gsmp}{\textsc{gsmp}\xspace}
\newcommand{\gspn}{\textsc{gspn}\xspace}
\newcommand{\filt}{\ensuremath{\mathcal{N}_{t\text{-}}}\xspace}

\title{Continuous-Time, Discrete-Event Simulation from Counting Processes}
\author{ANDREW J DOLGERT
\affil{Cornell University}}

\begin{abstract}
This is a method for discrete event simulation
specified by survival analysis. It presents
a sequence of steps. First,
hazard rates from survival analysis specify
the rates of a set of counting processes.
Second, those counting processes define a transition kernel.
Third, there are four different ways to sample
that transition kernel, including a first-principles
derivation of exact stochastic simulation algorithms
(\ssa) in continuous time. This simulation allows time-dependent
intensities which include both continuous and atomic components.
Separating the steps involved makes a clear
correspondence between mathematical formulation and
algorithmic implementation.
\end{abstract}


\acmformat{Andrew J.\ Dolgert. 2016. Continuous-Time, Discrete-Event Simulation from Counting Processes}

\begin{document}

\begin{bottomstuff}
This work was supported by USDA-APHIS Cooperative Agreement
No. 15-9200-0411-CA with Cornell University.

Author's address: A.\ J.\ Dolgert, (Current address)
Institute for Health Metrics and Evaluation, University of Washington.
\end{bottomstuff}

\maketitle

\section{Introduction}
A continuous-time, discrete-event (\ctde) simulation
can be based on any one of a number of stochastic processes,
including
Generalized Semi-Markov Processes (\gsmp)~\cite{Glynn1989},
Generalized Stochastic Petri Nets (\gspn)~\cite{Haas2002}, or chemical
stochastic processes~\cite{Gillespie2007} These processes vary in how they
define state and how they specify the distribution of times
at which each event will fire, but all of them
have the remarkable property that 
the hazard rate of any event in a realization of the process
will match the hazard rate of the distribution of times
used to specify that event.

In their work on chemical stochastic processes,
Anderson and Kurtz~\citeyear{anderson2015stochastic} offer
a construction of a stochastic process which can
serve as an alternative method of \ctde simulation.
Because it uses counting processes as the basis for the construction,
their work emphasizes the role of hazard rates in defining
the process. As a result, certain forms of \gsmp and \gspn, those
which avoid use of immediate events and simultaneous events, could be considered
subclasses for the purposes of computation.
The formalism of counting
processes also offers the algorithmic advantage of a clear separation
between defining a transition kernel and sampling that kernel.

Anderson and Kurtz present the more general stochastic process,
here called \emph{competing clocks}, as a step in derivation of
stochastic processes for chemical simulation.
Implications of the general stochastic process for
specification and computation are unexplored.
The goal of this article is to expand upon the competing clocks process
by clarifying how its specification relates to survival analysis
and to show how the machinery
of counting processes translates into modular and efficient computational
algorithms.
The scope of this article is time-homogeneous,
continuous-time simulation with discrete space because
discrete state spaces simplify the conditions under which a 
specification guarantees a single next time step.

In order to demonstrate the power of the counting process formalism,
these counting processes can have any regular distribution,
including those for which the hazard rate can have
both a continuous and atomic part. This will produce a novel
generalization of Next Reaction Method and Direct Method
exact stochastic simulation algorithms.
Johansen points out that atomic intensities can be useful
for data analysis, which makes them important for
simulation from such data analysis~\cite{johansen1983extension}.
Such estimates appear as Dirac functions in
published models, for instance in Viet et al~\citeyear{Viet2004}.
The Nelson-Aalen estimator of cumulative hazard rate is, itself,
discontinuous. Addressing distributions which can
contain discontinuities addresses simulations which
contain both purely continuous and purely discontinuous
distributions. Finally, the set of computable models
include all regular distributions, so they are included
for completeness.

Sec.~\ref{sec:competing-clocks} defines the stochastic process.
Sec.~\ref{sec:correspondence} connects that specification
with survival analysis.
Two consequences of this definition are
that implementation of a library to support these
simulations can mirror the mathematics quite
closely, Sec.~\ref{sec:computing}, and that
sampling techniques from
discrete-event simulation and chemical simulation
can be used together, Sec.~\ref{sec:sampling}.
Comparisons with previous work will be discussed in the conclusion.

\section{Competing Clocks Construction}\label{sec:competing-clocks}
Anderson and Kurtz present a stochastic process~\citeyear{anderson2015stochastic},
here called a
\emph{competing clocks process}.
This section extends that work only by including atomic hazards.
The stochastic process they derive as an intermediate
result towards applications in chemistry will be shown
in the next section to be
the basis for a continuous-time discrete event simulation.
This section has two steps, definition of a single \emph{clock process} as a counting
process with a time-dependent intensity, and then
composition of multiple clock processes into a competing clocks process.

The clock process is a counting process, $N(t)$, which starts at
$N(0)=0$ and is constant except of jumps of $dN=1$.
If the time of the $k$th jump is labeled $R_k$, the interarrival
time is $U_1=R_1$, $U_k=R_k-R_{k-1}$.
A counting process is characterized by its intensity, which is
a measure that assigns to every time a probability
the clock will jump in the next interval, $dt$. The hazard
rate is the probability of the \emph{next} jump, given the
condition that that next jump has not yet happened. The two
will have the same value over a time interval between jumps,
so they may be used interchangeably, but the intensity is
the intensity of the clock process while the hazard rate
is the hazard rate of the next jump.

A realization of a clock process is a complete set
of jump times.
At any time, $t$, the \emph{past} of the process is
a set of jump times for times less than $t$, and is called
a path, $v^1$, of the set of paths of $N$. The clock process,
as defined by Kurtz~\citeyear{Kurtz1980}, can also depend
on another path, $z$, of an external stochastic process,
$Z$, to represent the environment.
Both are restricted to have a limit
from the left and be continuous from the right, which is cadlag,
and the filtration of the clock process is the history
of both before a current time,
$\filt^1=\sigma(N(s), Z(s), s<t)$.

The intensity, then, is a function adapted to this
filtration which maps time and an element of $(v^1,z)\in(N,Z)$
to a non-negative number.
An intensity must be measurable and a.s.\ finite. It also
cannot depend on the future, so it depends on the paths $v^1$ and
$z$ only for times less than $t$.
The intensity has a continuous part which is
written as $\lambda(t,v^1,z)$ and an atomic
part, written using the Dirac delta function, $\delta$, as
\begin{equation}
  p(t,v^1,z)=\sum_i p_i(v^1,z) \delta(t-t_i(v^1,z)),\label{eqn:deltas}
\end{equation}
which implies each shock, $p_i(v^1,z)$, happens at time, $t_i(v^1,z)$.
Both continuous
and atomic parts of the intensity are predictable functions
of the filtration.

The clock process can be defined in terms of the given
intensity by stating that, for each jump time, $R_k$,
using the notation $t\wedge R_k=\mbox{min}(t,R_k)$,
\begin{equation}
M_k(t)=N(t\wedge R_k)-\int_0^{t\wedge R_k}\lambda(s,v^1,z)ds-\sum_{t_i<t\wedge R_k} p_i(v^1,z)
\end{equation}
is a martingale, such that $E[M_k(t+s)|\filt]=M_k(t)$.

Competing clocks are a set of clock processes, labeled
$j=1,2\ldots$.
For any single clock process, consider other clock processes
as its external environment, so that any one of them can depend
on a joint history of all clocks,
expressed as the filtration $\filt=\sigma(N_j(s), s<t,\forall j)$.
The paths $v$ are the paths of all of the set of clock processes.
Anderson and Kurtz permit the competing clocks to depend not
only on their joint history but also on an external environment,
but this presentation will exclude it to focus sampling algorithms
where the next stopping time can be determined entirely from
the hazards.

As for the single clock process, the
intensity must be a measurable function of the filtration,
and the sum of the intensities must be a.s.\ finite.
What links the clock processes together is that the intensity
of each clock process is a measurable function of the
paths of any of the clock processes. The system stopping
times, $T_n$ are the set of all jump times of all clocks, $R_k^j$.
No two clocks may jump at the
same time, which means that no atomic jumps $t_{ij}(v)$
may be the same among all of the atomic intensities,
\begin{equation}
  p_j(t,v)=\sum_i p_{ij}(v) \delta(t-t_{ij}(v)).
\end{equation}
The definition of the competing clocks process is an assertion
that
\begin{equation}
M_{nj}(t)=N_j(t\wedge T_n)-\int_0^{t\wedge T_n}\lambda_j(s,v)ds-\sum_{t_{ij}<t\wedge T_n} p_{ij}(v)\label{eqn:fullmartingale}
\end{equation}
is a martingale.
This concise definition determines stochastic processes
$N_j(t)$ from specifications for $\lambda_j$ and $p_j$. The next
section discusses how to choose those specifications.

\begin{table}
\begin{tabular}{lll}
Notation & Description \\ \hline
$N_j(t)$ & The $j$th lock process\\
$R^j_k$ & The time of each jump for a clock process \\
$\lambda_j(t,v)$ & Continuous intensity of clock process $j$ \\
$p_j(t,v)$ & Atomic intensity of clock process $j$ \\
$h_j(t)$ & Continuous hazard rate until the next jump of clock process $j$ \\
$a_j(t)$ & Atomic hazard rate until the next jump of clock process $j$ \\
$\nu_j$ & Jump mark for each clock process \\
$S_j(t)$ & Survival of clock process $j$ \\
$X(t)$ & Discrete state of the system \\
$T_n$ & Each stopping time of the system
\end{tabular}
\end{table}

\section{Correspondence Between a Model and a Process}\label{sec:correspondence}

Given a specific set of functions for the intensity,
Eq.~\ref{eqn:fullmartingale} implies a stochastic differential equation which
can then be solved as described in Sec.~\ref{sec:sampling}.
What choices of $\lambda_j(t,v)$ and $p_{j}(t,v)$ are meaningful representations
of disease progression, a financial system, or an influencer network?
A model, in this case, is a set of rules about what can change
next and the likelihood of a particular next change and time.
This breaks down into two questions, what can happen
and when it can happen.

The intensity of a clock process is written as a function
of two variables, $t$ and $v$. The second variable is
the set of jumps that have happened and times at which
they happened, so it changes only at stopping times
of the system. Assume $v$ is the history of the competing
clocks up to some time $T_n$.
The intensity can always be written as
two functions. The function $\phi_j(v)$ looks at $v$ to determine
whether this particular clock can fire and, if so,
at what rate. The functions $h_{jn}(t)$ and $a_{jn}(t)$ specify the
continuous and atomic hazard rate
for process $j$ given the past at time $T_n$,
\begin{eqnarray}
  \phi_j: v& \rightarrow &(h_{jn}(t), a_{jn}(t)) \label{eqn:hazardenabling} \\
  h_{jn}: t & \rightarrow & \lambda_j(t,v)\label{eqn:hazardstep} \\
  a_{jn} : t & \rightarrow & p_j(t,v).\label{eqn:atomicstep}
\end{eqnarray}
The first function, $\phi_j(v)$, determines what can happen.
If a clock process can fire at some time in the future,
according to the current $v$, then the hazard rate is
nonzero somewhere. If a clock process cannot fire,
then $\phi_j(v)$ maps to some other value, called
$\Delta$, which indicates a jump is forbidden. The function
$\phi_j(v)$ may return the same hazard rates for multiple stopping
times of the whole system, so that the functional form of the
hazard rates does not change. The most recent
time at which a clock became enabled is its
enabling time, $\mathcal{E}_j(v)$.
Specification of the intensity of a clock process is
specification of a sequence of hazard rates, each of which
is a hazard rate given a certain past.

One reason to state discrete event simulation as
competing clocks is to make an explicit connection to
hazards estimated by survival analysis.
The hazard rate in Eq.\ref{eqn:hazardstep} is for a single
stochastic process.
While this rate is applied to an individual clock process,
it usually comes from observation of an ensemble of
events. Aalen~\citeyear{aalen2008survival} describes
such survival analysis, which observes multiple
instances of an event and treats each such event, $C_j$,
with the same past, $\filt$,
as having the same hazard rate,
\begin{equation}
  \hat{h}(t)dt=P(t\le C_j<t+dt|\filt).
\end{equation}
A set of these events are grouped
together into a stochastic process that counts those not fired,
\begin{equation}
  Y(t)=\sum_j \mathcal{I}(C_j\le t),
\end{equation}
where the time is relative to a start time for each event,
its enabling time. From this is constructed the
intensity of a counting process that counts the number of events,
$\hat{N}(t)$, which happen before time $t$, 
\begin{equation}
  \hat{\lambda}(t)dt=\hat{h}(t)Y(t)dt.
\end{equation}
The integrated
intensity is the predictable part of the counting process
for the set of individual events,
\begin{equation}
  E\left[\left.\hat{N}(t)-\int_0^t\hat{h}(s)Y(s)ds\right|\filt\right]=0.\label{eqn:survivalmartingale}
\end{equation}
This is the statistical model used to estimate hazard
rates from observed outcomes. Survival analysis uses the aggregate of observed
event times to estimate the unknown hazard rate, while
the clock process uses known hazard rates to produce
sequences of event times, so they are almost the reverse of
each other, except that the Martingale in Eq.~\ref{eqn:survivalmartingale}
is for an ensemble of the same event with sufficiently similar past,
while the intensity
for a single clock process is for a single event.

Representations of a model as a competing clocks process
are not unique, but there is a most
expanded representation, where there is a separate
clock for each possible action in the system so that
each clock process can jump at most once.
For instance, $N$ individuals which could each make $k$ actions
would have $kN$ clocks. If an individual could repeat
actions, then there would be countably many clocks,
but only a finite number of them could be enabled
at any time. This would require that no jump by
an enabled clock could cause more than a finite number
of clocks to become enabled.

The hazard rates must match those estimated for the
real-world processes in order for the
competing clocks construction to be a faithful model.
This is different from saying the
specification must be explicitly in terms of hazard rates.
Sometimes the specified rate for a clock process may
be an Aalen-Johansen estimator, but it is more often
a parametric distribution, such as a Gamma distribution
or Weibull distribution, whose hazard rate matches
an estimate. The hazard rates themselves are not a
necessary computational tool for simulation. They are the
measure by which the simulation conforms to observations.

The intensity must be locally finite between jumps of the system
time $T_n$. The hazard rate could, for instance, be specified as a uniform
distribution between $T_a$ and $T_b$. The hazard rate
of the uniform distribution is infinite at $T_b$, but
the stochastic process is realized by sampling from the 
hazard rate, so the jump time will always be before $T_b$,
making the intensity locally finite.

\emph{Example: Rabbits Eating\/} Create a Poisson-distributed clock process,
$N_0(t)$, with rate $l$,
which represents a count of food produced. It is always enabled.
Each of $M$ rabbits has
a set $K_m$ of clock processes which represent varying
integral amounts of food, $d_k$, the rabbit might eat. 
A random variable representing uneaten food is
the total amount of food produced minus the counts
of all clock processes representing rabbit consumption,
\begin{equation}
  d=N_0(t)-\sum_{m=1}^{|M|}\sum_{k\in K_m} N_{mk}(t)d_k
\end{equation}
with multipliers to represent the amount consumed.
Each eating process, $N_{mk}(t)$, has an enabling rule, $d_k>d$,
so that one rabbit consuming food can disable 
eating processes for other rabbits at the jump time.
The enabling time of any eating process always
maps to the last time the rabbit ate as the maximum
jump time of the eating processes for that rabbit.
Call it $\mathcal{E}_m$.
Let the intensity of the eating clocks have a Weibull-distributed
interarrival time,
\begin{equation}
  S_j(t)=1-\exp((-(t-\mathcal{E}_m)/f)^2)
\end{equation}
where $f$ depends on the size of the last meal.

Each clock process has an enabling time which can
map to infinity if an enabling rule is not met, and it has
a hazard rate defined by supplying a known distribution.
The net result of these stochastic process is an
observable amount of uneaten food and rate at
which the rabbits eat when competing for the food.

\section{Computing a Realization}\label{sec:computing}
The competing clocks construction defines a model which
produces unique, computable realizations. This section
describes several arbitrary choices which are made for the sake
of efficiency of computation and algorithmic clarity.

\subsection{Natural algorithm to compute a realization}
\label{sec:natural-algorithm}
The minimum information needed to define a model is
how each jump affects the intensity for each
clock process. The filtration contains all of the
information required because it is the $\sigma$-field of the paths of each
of the clock processes. A cumulative intensity adapted
directly to the filtration is called \emph{natural}.

The algorithm to compute a realization will sample the next
jump, $j'$, and time, $T_{n}$, given the filtration at the
previous jump, $T_{n-1}$. In terms of the minimum information,
there are three steps to obtaining the next state and
time of the stochastic process, given the past.
\begin{enumerate}
  \item Compute the transition kernel. Sec.~\ref{sec:sampling}
  will describe two versions of a transition kernel, both of
  which depend on the value of the intensity as a function
  of a current time and system path,
  \begin{equation}
    (T_{n-1}, v)\rightarrow(\lambda_j(t,v), p_j(t,v)).
  \end{equation}
  \item Sample the jump time and jump mark for the next transition.
  \begin{equation}
    (\lambda_j(t,v), p_j(t,v))\rightarrow (j', T_{n})
  \end{equation}
  \item Apply the transition function to update the path of the system.
  \begin{equation}
    (T_{n-1}, T_{n},v, j')\rightarrow (T_{n},  v')
  \end{equation}
  Changing the filtration implies updating any random variables
  adapted to that filtration, such as cumulative intensities.
\end{enumerate}
The natural algorithm does not need reference to
a system state.

\subsection{Discrete State}
The state of the system is a random variable derived
from the self-exciting filtration. As described by
Anderson and Kurtz~\citeyear{anderson2015stochastic},
a compact definition of a discrete
state is to associate with each clock process
a vector $\nu_j$ which increments an initial state
$X(0)$ as
\begin{equation}
  X(t)=X(0)+\sum_{j}N_j(t)\nu_j.
\end{equation}
The vectors, $\nu_j$, can be considered jump marks for the
clock processes.
There is no restriction that the vectors $\nu_j$ be unique.

In many cases, every intensity in the system is a function
of the history only through the state. In the broadest case,
every intensity could depend on the full path of the state,
its value at every time in the past, as $\lambda_j(t,X(t))$.
Other systems restrict the intensities to depend only on
the most recent value of the state, some $X(T_k)$, and the time at which
the state entered that most recent value, $\lambda_j(t, X(T_k), T_k)$.
These kinds of restricted forms of the intensity depend
on the structure of the analysis which produced the
specifications for hazard rates. Each such restriction
can reduce the amount of computational state required for
a sampling algorithm.

\subsection{Dependency graph}
The intensity of any clock process will be a function
of the paths of some subset of clock processes. This
creates a directed graph from a clock process to those
clock process whose jumps may affect its intensity.
The dependency graph provides
an efficient way to calculate which clocks' intensities
need to be re-evaluated after any clock fires.
For some models, such as a susceptible-infectious-recovered
(\textsc{sir}) model~\cite{allen1994some},
a natural algorithm with a dependency
graph may be the most efficient expression of the model.

Gibson and Bruck observed~\citeyear{Gibson2000} that if the form of the intensity is such that it is a function
of the state, for instance $\lambda_j(t,X(T_k))$,
then the dependency graph may be written as a directed,
bipartite graph. The two
types of nodes in the graph are substates of the state,
$X=(x_1, x_2, x_3\ldots)$, and the
clocks $N=(N_1, N_2, N_3\ldots)$. There is an edge
from every substate to all clocks whose jumps would change
that substate. There is an edge from 
every clock to all substates on whose path it depends. In most cases, a bipartite
representation of the dependency graph will require
less memory than a dependency graph among clock processes.

The dependency graph also serves as a form of parameterization
of the clocks. When multiple individuals can have similar
behavior, there will be cases where two clock intensities can be written
as the same functional form applied to different substates,
\begin{eqnarray}
\lambda_1(t,X(t))&=&l(t,x_1,x_2) \\
\lambda_2(t,X(t))&=&l(t,x_3,x_4).
\end{eqnarray}
In this case, the specification of the intensity can be
the common function, $l$, and projections of the state,
defined by the dependency graph.

The reaction graph used in chemical simulation
makes assumptions about the dependency graph and encodes
more information. Reaction graphs annotate transitions,
equivalent to clock processes, with rates and annotate edges
with stoichiometric constants which are equivalent to the $\nu_j$.
For chemical simulation, knowing the stoichiometric constants
of a transition
also defines enabling rules and parameterizes intensity functions.

In order for one clock jump to be the cause of
another, there must be a path from one to the other in
the dependency graph. It is the topology upon which
the intensities define the dynamics.
The dependency graph encodes much of the order
of computation as a result. The number of nodes,
the number of clocks, and the distribution of cardinalities
of the clocks and nodes determine the
number of instructions necessary to sample the
next jump of a realization.

\subsection{Finiteness}
There are some expected requirements for finiteness.
The number of enabled processes at any stopping time
must be finite, both for representation on a computer
and because the sampling algorithms in Sec.~\ref{sec:sampling} rely on finite
alternatives. The number of edges in the adjacency graph
from one clock to substates and to other clocks must
be finite, as well, because they will be checked
for enabled clocks at each jump of the system.

More interesting is what need not be finite. The space
of substates could be infinite. A common representation
for the state is a map from an index to the substate,
and setting a state to zero can be done by removing 
the substate from the map. In this way, a simulation
of an ant could allow it to walk an infinite plane.
Similarly, a stochastic process could be defined for
an infinite number of clocks for which only a finite
number are enabled at any time.

\subsection{Stateful clock hazard rates}
An important optimization is saving the hazard rates
of clocks between jumps of the system. The intensity function
was described in Eqs.~\ref{eqn:hazardenabling}--\ref{eqn:atomicstep} as having two parts, one which
calculates a clock's current hazard rate and the clock's
hazard rate which is a given function of time.
The set of current hazard rates, $h_{jn}(t)$ and $p_{jn}(t)$, can
be kept as computational state between jumps
as long as that set is updated to account for
additions, subtractions, or modifications to clock hazard rates.
The hazard rates are not part of the minimum state of the
stochastic equation because they are predictable
functions of the history, $v$ or $X(t)$.

\subsection{Computing a realization as a finite automaton}
\begin{figure}[t]
\centerline{\includegraphics[scale=0.4]{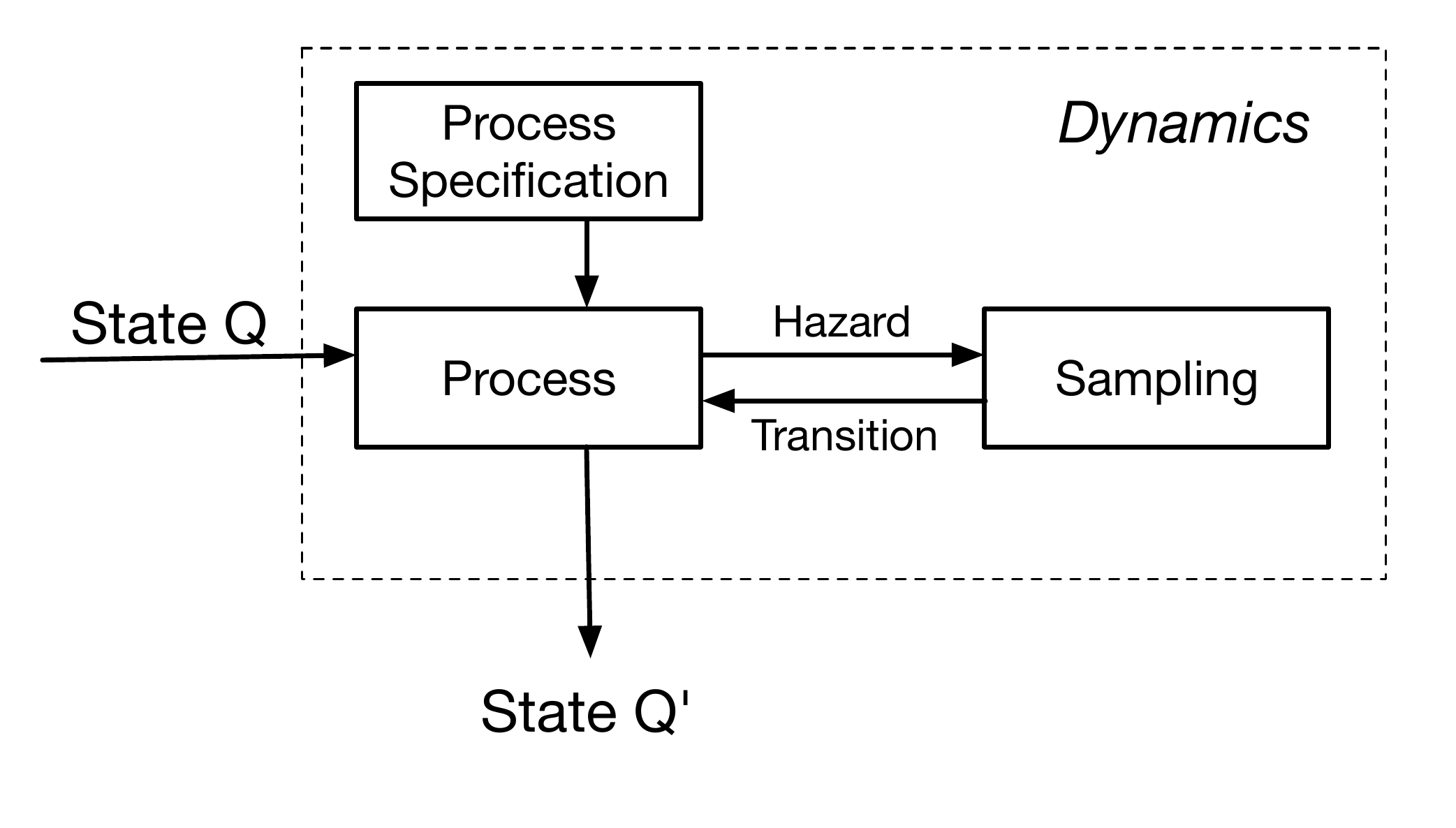}}
\caption{The core of a finite automaton to calculate a trajectory
depends upon both the statistical process and its sampler.\label{fig:automaton}}
\end{figure}%
This section organizes computation of a realization into the
form of a stochastic finite automaton with no inputs and
nondeterministic output.
A sequential machine~\cite{arbib1980machines} is a quintuple $(X_0, Q, \delta, Y, \lambda)$,
where $X_0$ is the input set, $Q$ the set of states, $\delta$ the
dynamics, $Y$ the set of outputs, and $\lambda$ the output
function. (The notation, $\lambda_j$, will refer below solely to clock
intensities.) This section will describe the state and dynamics
of that sequential machine.

The dynamics
are a function from the state and input set to a new state,
where the state, in this case, refers to all mutable state
of the system, which is the discrete state $X(t)$,
the current hazard rates, $h_{jn}(T_n)$,
and any state associated with sampling.
The entities that define the dynamics are the space of
the discrete states, the dependency graph, and the clock
intensities.
For each clock, specification includes an enabling
function which returns also whether enabling has
changed since the last discrete state, and, for each
enabled clock, a function
which returns its current hazard rate.

The three steps in Sec.~\ref{sec:natural-algorithm}
are the dynamics of the finite automaton. Written in terms
of discrete state and caching of hazards, they become the following:
\begin{enumerate}
  \item Compute the transition kernel.
  \begin{equation}
    (T_{n-1}, X(T_{n-1}))\rightarrow
    (h_{j,n-1}(t,X(T_{n-1})),a_{j,n-1}(t,X(T_{n-1})))
  \end{equation}
\item Sample the jump time and jump mark for the next transition.
  \begin{equation}
    (h_{j,n-1}(t,X(T_{n-1})),a_{j,n-1}(t,X(T_{n-1})))\rightarrow (j', T_n)
  \end{equation}
  \item Apply the transition function to update the state.
  \begin{equation}
    (T_{n-1}, X(T_{n-1}), \nu_j, j', T_{n})\rightarrow (T_{n}, X(T_{n}), h_{j,n}(t,X(T_n)), a_{j,n}(t,X(T_n)))
  \end{equation}
  Applying the transition function requires using the dependency graph to
  update all hazard rates which change. During that update, the hazard
  rates, as a group, do not reflect the state of the system.
  Any derivative quantities, such as integrated hazard rates,
  need to be updated during this invariant-breaking step.
\end{enumerate}
Output is separate from the dynamics and is a function
of the filtration.
The output observable of the sequential machine could be
the filtration directly, $(T_n, j)$, or the set of
states and times, $(T_n, X(T_n))$, or any random variable computed
from these.

\section{Sampling}\label{sec:sampling}

\subsection{First Reaction Method}
The martingale definition of the competing clocks
process, Eq.~\ref{eqn:fullmartingale}, is equivalent
to a stochastic equation whose solution defines
the transition kernel to sample. Take any of the clock processes,
and label the survival of its interarrival time since
its last jump at $R_{k-1}$ as $S_j(t)=P[R_k-R_{k-1}>t]$.
Write the cumulative intensity as
\begin{equation}
\Lambda_j(t)=\int_0^{t}\lambda_j(s,v)ds+\sum_{t_{ij}<t} p_{ij}(v),
\end{equation}
where $v$ is the path for all of the clock processes.
The probability
definitions for the survival and hazard relate the two as
\begin{equation}
  dS_j(t)= -S_j(t^-)d\Lambda_j(t),\label{eqn:differential-survival}
\end{equation}
where $S_j(t^-)$ indicates the value of the survival just
before time $t$. Eq.~\ref{eqn:differential-survival} is a
Dol\'eans-Dade differential relation whose solution is given by product
integration~\cite{Gill1990},
\begin{equation}
  S_j(t)=e^{-\int_{R_{k-1}}^t\lambda_j(s,v)ds}
  \prod_{R_{k-1}< t_i\le t}(1-p_{ij}(v)).\label{eqn:singlesurvivalputatitive}
\end{equation}
These are the distributions of jump times for each clock process to fire,
were it the first to fire next. As such, they are called putative distributions.

The name \emph{competing processes} comes from Markov renewal processes,
where, at any time step $T_{n-1}$, the next jump of the process can be
sampled by transforming the problem into $j$ separate distributions
in time~\cite{Howard2007}.
The First Reaction Method for competing clocks substitutes, at a time $T_{n-1}$,
the current hazard rate for the intensity, using
$\phi_j(v)\mapsto (h_{j,n-1}(t),a_{j,n-1}(t))$. It then samples the distributions
defined by these hazard rates to determine which clock process has the soonest
jump time. Given independent uniform random numbers, $U_j\in[0,1)$, and
cumulative distributions, $F_j(t)=1-S_j(t)$, the sampling algorithm is
\begin{equation}
  t_j' = F_j^{-1}(U_j)\qquad T_n=\min(t_j')
\end{equation}
The algorithm chooses the minimum jump, updates intensities, and generates
a fresh set of random numbers and putative firing times, $t_j'$.
Gillespie named this technique the First Reaction Method~\citeyear{Gillespie1978},
and it differs from classic competing processes in that more than one of the
jumps may cause the system to enter the same state.

\subsection{Next Reaction Method}
Every sampling method for competing clocks samples a joint
distribution over which clock jumps and when is the next jump.
While the First Reaction Method samples at each time interval
separately, the Next Reaction Method~\cite{Gibson2000}
examines each clock process separately, saving samples between
times $T_n$. Justification
for saving samples rests on proving that competing clocks processes
are independent, doubly-stochastic Poisson processes.

Unit Poisson processes have the special property that repeated sampling
for the next to fire from a set of Poisson processes can save the
times of all samples that were later than the soonest~\cite{van1992stochastic}.
Denote unit Poisson processes by $N_{\pi j}(\tau)$ and their cumulative distributions by
$F_{\pi j}(\tau_j)=1-e^{-\tau_j}$. Generate putative firing times,
$\tau_j'=F_{\pi j}^{-1}(U_j)$, where $U_j\in[0,1)$. After selecting the next time as $T_n=\mbox{min}(\tau_j')$
and removing it from the set of times, subtract $T_n$ from the remaining
$\tau_j'$. The distribution of $\tau_j'$ will still be exponentially-distributed
and can be used to choose the next soonest time. The Next
Reaction Method applies this same technique to a situation where
each Poisson process has a clock which is itself a stochastic variable.

Kallenberg~\citeyear{Kallenberg1990} showed that it is possible to write a
clock process as a unit Poisson process which 
depends on a time process,
\begin{equation}
  L_j(t,v)=\int_{0}^t\lambda_j(s,v)ds
  +\sum_{0< t_i\le t}\ln(1-p_{ij}(v)).
\end{equation}
Then the clock process 
is explicitly doubly-stochastic,
\begin{equation}
N_j(t,v)=N_{\pi j}(L_j(t,v)).\label{eqn:doubly-stochastic}
\end{equation}
One way to see that Eqs.~\ref{eqn:doubly-stochastic} and~\ref{eqn:singlesurvivalputatitive} are
equivalent is to compute Eq.~\ref{eqn:differential-survival}
for the clock process,
\begin{equation}
  dS_{\pi j}(L_j(t,v))=-S_{\pi j}(L_j(t,v))d\Lambda_j(t,v),
\end{equation}
which means that $S_{\pi j}(L(t,v))$ has the same
infinitesimal generator as $N_j(t,v)$, so it has the same survival.
When the intensity is purely continuous, the time
process coincides with the integrated intensity.
In the more general case including atomic
contributions, the integrated intensity is different from the time process.

Aalen and Hoem proved that sampling doubly-stochastic processes
with continuous hazards
doesn't introduce bias, despite explicit dependence on each other
through $v$~\citeyear{Aalen1978}.
Extension of independence among the unit Poisson
processes to time processes of discontinuous cumulative intensity was
remarked to be an open question~\cite{Vere-Jones2004}.
As is sometimes the case, the computational framework
proposed here will rest upon a result which is
reasonable and seen to work although it is unproven.

Saving the $\tau_j'$ from one step to the next is equivalent to
saving $S_j'=1-U_j=\exp(-\tau_j')$ from one step to the next.
Each clock process, from the start of the simulation, has a sampled
survival, $S_j'$, and, whenever that clock process is enabled, it consumes
from that survival at a rate given by $L_j(t,v)$ until it fires.
Aalen and Hoem's work gives permission to save $S_j'$ from
one jump to the next, \emph{as long as each step removes and resamples
the $S_j'$ of only the process that jumps.}
All variations on the Next Reaction Method are different
ways to maintain the set of survivals given the need, at each time step,
to find the minimum of the $t_j'$ and then update the time processes
when $v$ now includes a new jump and time.

Gibson and Bruck's algorithm labels with $t_i$ those
stopping times when $\phi_j(z)$ changes for some clock process $j$~\citeyear{Gibson2000}. The conditional
survival is the survival to some $t_2$ given that the
process survived to $t_1$, and it is multiplicative,
so that $S_j(t)=S_{j0}(\mathcal{E}_j(v),t_1)S_{j1}(t_1,t)$
if the process survived past a time $t_1$.
When a clock process is first enabled, sampling
by inversion solves
\begin{equation}
  S_j'=S_{j0}(\mathcal{E}_j(v),t')
\end{equation}
for $t'$.
If the functional form changes at a stopping time
$t_1$ before the clock has jumped, the
Next Reaction Method rewrites the original inversion
equation as
\begin{equation}
  S_j'=S_{j0}(\mathcal{E}_j(v),t_1)S_{j1}(t_1,t'),
\end{equation}
which uses the property that the survival is multiplicative.
This is called shifting the distribution.
Labeling the enabling time as $t_0=\mathcal{E}_j(v)$, the general form is
\begin{equation}
  S_j'=S_{jn}(t_n,t')\prod_{m=0}^{n-1}S_{jm}(t_m,t_{m+1})\label{eqn:nr-invert}
\end{equation}
Gibson and Bruck's version of the Next Reaction Method
maintains the value of
\begin{equation}
  S_{jn}'=S_j'/\prod_{m=0}^{n-1}S_{jm}(t_m,t_{m+1})
\end{equation}
as computational state with which to solve $S_{jn}'=S_{jn}(t_n,t')$.

The survival can also be written in terms of the
time process, and, working with a continuous form of the time
process, Anderson showed~\citeyear{Anderson2007} that its conditional
form
\begin{equation}
  L_j(t_1,t_2,v)=\int_{t_1}^{t_2}h_j(s)ds
  +\sum_{t_1< t_i\le t_2}\ln(1-a_{ij}).
\end{equation}
will express the Next Reaction Method as an additive
rule,
\begin{equation}
  \ln(S_j')=L_{jn}(t_n,t_j',v)+\sum_{m=0}^{n-1}L_{jm}(t_m,t_{m+1},v).\label{eqn:and-invert}
\end{equation}
The running value of the second term is saved as $L_{jn}'$.
For some distributions, such as the exponential and
Weibull distributions, this algorithm is involves only
addition instead of exponentials, making it preferable
to Gibson and Bruck's formulation.

The algorithm for the Next Reaction Method can choose whether
to update resampled jump times using the survival, Eq.~\ref{eqn:nr-invert},
or the time process, Eq.~\ref{eqn:and-invert},
depending on which is most efficient for the distribution
of a particular clock's next jump. Any distribution used in a simulation
needs to support three operations.
\begin{enumerate}
  \item The first sample need not use inversion, but it must
  return the survival of the draw,
  \begin{equation}
    S_{j0}(t), \mbox{rng}\rightarrow (t_j', S_j')
  \end{equation}
  or the log of the survival,
  \begin{equation}
    S_{j0}(t), \mbox{rng}\rightarrow (t_j', \ln(S_j')),
  \end{equation}
  depending on whether conditional survival or
  time process will be used.
  \item Later samples must match the survival of the initial
  draw, usually using sampling by inversion,
  \begin{equation}
    (S_{jn}(t_n,t), S_{jn}')\rightarrow t_j'\ \mbox{or}\ (L_{jn}(t_n,t), \ln(S_{jn}'))\rightarrow t_j'
  \end{equation}
  Distributions without an analytic inverse require numerical inversion.
  \item Each distribution must also calculate the cumulative survival or cumulative time process, given a time interval over which the associated clock has
  not fired,
  \begin{equation}
    (S_{j,n}(t_n,t), T_n)\rightarrow S_{jn}'\ \mbox{or}\ 
    (L_{j,n}(t_n,t), T_n)\rightarrow L_{jn}'.
  \end{equation}
\end{enumerate}
By exploiting the similarity between the
survival and hazard techniques, it is possible to create
a single interface which permits sampling any distribution
with the most appropriate technique.

\subsection{Next to Fire}
Part of the definition of the Generalized Semi-Markov
Process (\gsmp) is how to sample it~\cite{Glynn1989}. The method is to
sample all enabled clocks, select the soonest, and then
resample those which were affected by the soonest.
If a clock is non-exponential, or if its exponential rate
has changed, then resample the clock with a shifted
distribution, just as in the Next Reaction Method.
What differs from the Next Reaction Method is that there
is no effort to resample using the same cumulative survival.

The chief disadvantage is that resampling requires generation
of a new random number, which is now a relatively less expensive
operation than it once was. The great advantage for computational
efficiency is that there is more choice in how to sample
given distributions.
There are stability advantages
to sampling by inversion, but using sampling methods
specific to distributions can be much faster.
Sampling non-exponential continuous distributions requires care. A panoply
of techniques are covered in Devroye~\citeyear{devroye}, and methods for automatic
generation are discussed by H\"ormann et al. who have written
a software library,
\textsc{unu.ran}, which  which calculates
distributions using several methods for inversion while
providing calculation of the conditional survival~\cite{hormann2000random}.

For sampling, distributions
which are are both continuous and atomic
form a mixture model. Sampling continuous distributions
by composition techniques breaks the problem into a discrete
choice over mixture components and then continuous sampling
within the component.

While numerical inversion and integration of distributions
can be time-consuming, the central object of Next to Fire algorithms is
the calculation of the clock process whose next jump time
is soonest. The dependency graph controls which jump times
are added, removed, or modified in a set of putative jump times.
Using this graph to speed updates is the second contribution
to computational efficiency.
The computer science name for this algorithm is an
indexed arg min, or an indexed arg max with an objective function
which is $-t_j$. The most common data structure for storing
putative times is an indexed heap. While the Fibonacci
heap has the best order of computation for large simulations,
pairing heaps have better performance in most cases and
are significantly simpler to implement correctly~\cite{Mauch2011}.

\subsection{Direct Methods}
The previous three sampling techniques took independent
samples of the clock processes. The class of direct methods
constructs the joint density of the next clock to jump
and time at which it jumps in order to make only
two samples at each step of a realization.

The result for purely continuous
hazard rates is well-known, and the result can be obtained
in general from Jacod's theorem~\cite{jacod1975multivariate}.
Appendix~\ref{sec:derive-competing-risks} 
derives competing risks for discontinuous cumulative hazards.
The survival in the current state is
\begin{equation}
  P[T_n>t]=\exp\left[{-\int_{T_{n-1}}^t\sum_j h_{j}(s)ds}\right]
  \prod_{T_{n-1}<t_j\le t, \forall j}(1-p_{0j}(t)).\label{eqn:recap-cif-survival}
\end{equation}
and the cumulative incidence function is
\begin{equation}
  P[T_n\le t, J=h]=\int_{T_{n-1}}^t P[T_n>s]h_{j}(s)ds +
  \sum_{T_{n-1}<t_j\le t, \forall j}P[T_n>t-]p_{j}(u).\label{eqn:recap-cif-mixed}
\end{equation}
The cumulative incidence function is a joint distribution
over the discrete set of clock processes and the continuous
time for the next event. The direct method samples this joint distribution
using the method
of conditional probability, which takes the density,
separates it into marginal and conditional, and samples
first the marginal, then the conditional.

The density of Eq.~\ref{eqn:recap-cif-mixed} is called the
the cause-specific density,
\begin{equation}
  dQ(j,t)=P[T_n>t]h_{j}(t)dt + P[T_n>t-]p_{j}(t).
\end{equation}
The \emph{waiting time factorization} marginalizes over
the $j$ to produce a waiting time, $dW(t)=\sum_jdQ(j,t)$, and
a conditional stochastic matrix, $\pi_{j}(t)$.
Algorithms which sample the waiting time factorization
are called \emph{direct methods}, and they proceed in
two steps. First sample the waiting time density,
\begin{equation}
  U_1'=\int_0^{t'} dW(s)=P[T_n<t']\label{eqn:waiting-sample}
\end{equation}
and then sample the stochastic matrix given that time
for the smallest $j'$ such that
\begin{equation}
  U_2'\le  \sum_{j=1}^{j'} \pi_j(t').
\end{equation}
Sampling for $\tau'$ tells us will either return a unique
atomic hazard, which immediately tells us which clock jumped,
or it returns a continuous hazard, for which
\begin{equation}
  \pi_j(t')=\frac{h_j(t')}{\sum_kh_k(t')}\label{eqn:stochastic-inversion}
\end{equation}
permits a discrete draw over the clocks.

Given that the cause-specific density has two variables,
there is a second factorization, rarely used for computation,
called the \emph{holding time factorization}. It entails
calculating the marginal over times,
\begin{equation}
\pi_j=\int_0^\infty dF=P[T_n<\infty,J=j),
\end{equation}
which is just the stochastic matrix. The holding time
is a probability for a jump to a state given that
the state is already decided, $dF(j,t)/\pi_{j}$.
For exponential distributions, the holding time and waiting
time factorizations are indistinguishable.

Stating the waiting time as the sample of a univariate
distribution, as in Eq.~\ref{eqn:waiting-sample},
reduces computation to a standard set of techniques
covered by sampling theory, with the additional optimization that the
set of hazards remain consistent from jump to jump, except
as enabled, disabled, or modified according to the
dependency graph.

Were all of the hazards constant, the problem reduces to 
calculation of the sum $\sum_j h_{j}(t)$ given updates to
the list of enabled hazards. Further, Eq.~\ref{eqn:stochastic-inversion} relies on partial sums of hazards. Algorithms
to maintain partial sums of a mutable list are called
\emph{prefix sum algorithms} and are covered nicely
by Blelloch~\cite{blelloch1990prefix}. For instance, a binary heap can quickly
calculate prefix sums for a system with a constant total
number of enabled clocks. While different methods for
the partial sum algorithm have spawned different names
within the chemical simulation community, they represent
a well-contained point of variation for an implementation,
and the range of choices are explored in an
encyclopedic paper~\cite{Mauch2011}.

For non-exponential continuous hazards, the Eq.~\ref{eqn:waiting-sample}
requires inversion of the sum of hazard functions.
For certain distributions, such as Weibull, this inversion
may be analytic. In most cases, it is not. Direct
integration may be practical for small problems on many-core
architectures, but rejection-based algorithms can avoid
integration. Thanh and Priami use a composition-rejection
algorithm for hazards of chemical processes, constructing
a piecewise-linear envelope~\cite{Thanh2015}, and Holubec
applies what Devroye calls the thinning method, which is exact
for bounded hazards~\cite{Holubec2011}.

\subsection{Hierarchical}\label{sec:hierarchical}
Given the possibility that some clock processes are exponentially-distributed,
some atomic, and some non-exponential, all within the same
simulation, hierarchical sampling methods can be important
for efficiency. There are two different kinds of hierarchical
methods for exact simulation and the choice among these
methods depends on model size, the structure of the dependency
graph, and types of distributions for sets
of clocks to most appropriate sampling algorithms~\cite{Mauch2011}.

Direct methods which sample the waiting time factorization
must sample by inversion a sum of hazard rates. Because
a sum of sums is itself a sum, it is common to group
hazards according to special properties which make sums
easier to compute. For instance, a set of constant hazards,
all with the same value, can be represented by a binary
tree over integers. There may be sets of hazards which
are updated together, amortizing time for modification
of cumulative sums.

Similarly, sampling independent Poisson processes by the
Next Reaction Method or the Next to Fire method finds
the soonest to jump among all clocks. Some subset of clocks
can use a direct method. Others can use a first reaction
method. Then the minimum time of all putative times
is the next jump.

\section{Conclusion}

Competing clocks model a broad class of simulations.
They can be understood by comparison with the two most similar
techniques, chemical simulation and the \gspn.

Chemical simulation in continuous time has a discrete space
of reactants, which are chemical
components~\cite{Gillespie2007,Gibson2000}. It
establishes a separate stochastic
process in continuous time for each possible reaction
of those reactants, and the reaction rates are hazard
rates, called propensities. They depend on the current
state, $X(t)$. While competing clocks are defined
on jump times, with the state as a predictable variable,
the state of the system in chemical simulation is the state of the
reactants and interarrival times of the reactions.
Time-dependence in hazard rates tends to come from
time-inhomogeneous constraints, such as changing the volume
of a container. How an intensity might depend on
the path of $X(t)$ is less relevant for this physical system.
Chemical simulation can make a set of assumptions which
are a special case of competing clocks.

The \gspn is built around the transition from a state $s$
to a new state $s'$, whereas the insight of Anderson and
Kurtz is to build competing clocks from the path
of events and how it defines a transition kernel.
Built into the definition of the \gspn is
how to sample clock readings, whereas that is a distinct
question for competing clocks. Nevertheless, likelihood
of a realization of both processes, given the same
specification, takes the same form. In general, the \gspn
is a more thorough tool for engineering simulation
because techniques for immediate events and simultaneous
events are appropriate for models of management and
human-designed automation.

The clock-setting distribution of a \gspn
is the distribution generated by the hazard rate of that
event. It is defined as a function of four quantities
for the \gspn, the previous state, the most recent state
which enabled the distribution, the event that caused
that state, and the event that is enabled. For competing
clocks, the hazard rate can depend on any part of the
filtration, or variable derived from the filtration,
since the last time that clock fired, because each
clock process is regenerative. If a clock just fired
and is being reenabled, then it cannot depend on anything
but the current state. On the other hand, if a clock
has been disabled for some time since it last fired, or
since the simulation began, then it can depend on any
part of the filtration since that time.

In contrast to both chemical simulation and the \gspn,
competing clocks have no useful mechanism for specification
of which clock is enabled or disabled after a transition.
That decision is made by a specified function, $\phi_j(v)$, for
which the other two offer a more structured approach. This does allow, however,
for other ways to specify the causal chain of what
enables and disables events.

Writing a simulation with competing clocks is remarkably
direct. The state, if there is one, must be a map from a key to a state value.
Each clock process is defined by three things. The
enabling function looks at the state and says yes or no.
There can be some stateful memory because the intensity
can depend on the path of the system, so the intensity function
looks at both the current state and whatever it has
saved from past observations in order to generate
a distribution for when the clock will putatively jump.
The firing function of the intensity is just a vector of
how it changes the discrete state. If a dependency graph is
used, then these functions can be parameterized by
which states neighbor the clock process on the dependency graph.

Sampling of the process is separate from statement of a
particular clock process, much like in chemical simulation.
As a result, it is possible to build a library with
which to build optimized simulations of clock processes.
The authors have contributed such a library, written in
the Julia language, using \textsc{unu.ran} for generation of random
variates~\cite{dolgert2016}.

Extensions to the clock processes, as presented here, would
not hinder the connection to survival analysis. For instance,
Aalen discusses counting processes on continuous state
spaces~\cite{aalen2008survival}. This article excluded
the use of a process, $Z$, to represent the external environment
because it introduces the complication, in Eq.~\ref{eqn:hazardenabling},
that $\phi_j(v,z)$ would need to allow for the path, $z$,
to modify its output hazard rate at times \emph{before}
the next stopping time sampled by the kernel. If the external
environment is treated as a known quantity before the simulation,
its inclusion is trivial. If it is treated as a stochastic input,
it changes sampling methods significantly.

Also, as mentioned earlier, instantaneous
transitions are a known convenience. In addition,
the jump of a single clock could be the marginal sample
for a conditional sample among multiple outcomes. A hierarchical
sampling strategy can reduce the order of computation
significantly in some cases.

\appendix
\section*{APPENDIX: Derivation of Competing Risks for Discontinuous Hazards}
\setcounter{section}{1}
\label{sec:derive-competing-risks}
Aalen derives competing risks in continuous time using
a product integral formulation~\cite{aalen2008survival}.
This derivation follows the same method using regular distributions.
It is modeled by constructing
the transition matrix, $\mathbf{P}(t|s)$ for times $s$ and $t$, in terms of the
cumulative hazard matrix, $\mathbf{H}(t)$. Consider a system with
an initial state $0$ and final states $(j=1,2,3,\cdots)$.
We will write equations as though there were two final
states, but it generalizes to countably many.
The hazard matrix for the continuous part is
\begin{equation}
dH_c=
\begin{array}{ccc}
-\sum h_{j}(t) & h_{1}(t) & h_{2}(t)\\
0 & 1 & 0 \\
0 & 0 & 1
\end{array}.
\end{equation}
The hazard matrix for the atomic part is
nonzero only when one of the atomic jumps is nonzero,
so it takes the form of delta functions as in Eq.~\ref{eqn:deltas},
\begin{equation}
dH_a=
\begin{array}{ccc}
-\sum p_{j}(t) & p_{1}(t) & p_{2}(t)\\
0 & 1 & 0 \\
0 & 0 & 1
\end{array}.
\end{equation}
In product integral form, this matrix equation for the
survival is
\begin{equation}
  \mathbf{P}(t|s)=\Prodi_s^t(1-d\mathbf{H}),
\end{equation}
which tells us that the only nonzero components of the
transition matrix are in the first row,
\begin{equation}
P(t|s)=
\begin{array}{ccc}
P_{0}(t|s) & P_{1}(t|s) & P_{2}(t|s) \\
0 & 1 & 0 \\
0 & 0 & 1
\end{array},
\end{equation}
so we can solve for the transition matrix in two steps,
avoiding direct calculation of the product integral.
The survival in the current state, $P_{0}(t)=P[T>\tau]$,
can be formulated as a univariate transition whose cumulative
intensity is the sum of each of the cause-specific intensities.
Its value is
\begin{equation}
  P_{0}(t|s)=e^{-\int_s^t\sum_j h_{j}(s)ds}\prod_{s<t_{j}\le t, \forall j}(1-p_{j}(t)).\label{eqn:cif-survival}
\end{equation}
For transition to any other state, the Kolmogorov forward equation
gives the transition matrix in terms of the survival and the hazard,
\begin{equation}
dP_{j}(t|s)=-P_{0}(t-)dH_{j}(t)
\end{equation}
which makes the \emph{cumulative incidence function}, $P[T\le \tau, J=h]$
\begin{equation}
  P_{j}(t|s)=\int_s^t P_{0}(s)h_{j}(s)ds + \sum_{s<t_j\le t}P_{0}(t-)p_{j}(t).\label{eqn:cif-mixed}
\end{equation}
For Markov systems, the cumulative incidence function is called the
semi-Markov matrix.
Note that the survival appears as a limit from below so that
the atomic contribution at $t$ is not included in $P_{0}(t-)$.

\begin{acks}
The author would like to thank David Schneider and Chris Myers
of Cornell University for advice and support.
\end{acks}

\bibliographystyle{ACM-Reference-Format-Journals}
\bibliography{timedep}
\received{Month Year}{Month Year}{Month Year}
\end{document}